\documentclass[aps, prd, twocolumn, lengthcheck, superscriptaddress, showpacs, letterpaper, nofootinbib]{revtex4-1}

\usepackage{amsmath,amssymb}
\usepackage{graphicx,bm}
\usepackage{epstopdf}
\usepackage{ulem}
\usepackage[usenames]{color}
\usepackage{float}
\usepackage{multirow}
\usepackage{slashbox}

\def\la{\langle}\def\ra{\rangle}
\def\be{\begin{eqnarray}}\def\ee{\end{eqnarray}}
\def\lsim{\mathrel{\rlap{\lower3pt\hbox{\hskip1pt$\sim$}}
     \raise1pt\hbox{$<$}}} 
\def\gsim{\mathrel{\rlap{\lower3pt\hbox{\hskip1pt$\sim$}}
     \raise1pt\hbox{$>$}}} 
\def\le{ \begin{array}{ll}}\def\re{\end{array}}

\def\lear{ \left( \begin{array}{cc}}\def\rear{\end{array} \right)}

\def\le{ \left( \begin{array}{cc}}\def\re{\end{array} \right)}

\def\bi{\bibitem}

\begin{document}

\title{Scale-chiral symmetry, $\omega$ meson and dense baryonic matter}

\author{Yong-Liang Ma}
\email{yongliangma@jlu.edu.cn}
\affiliation{Center for Theoretical Physics and College of Physics, Jilin University, Changchun, 130012, China}

\author{Mannque Rho}
\email{mannque.rho@cea.fr}
\affiliation{ Institute de Physique Th\'eorique, CEA Saclay, 91191 Gif-sur-Yvette c\'edex, France }

\date{\today}

\begin{abstract}
It is shown that explicitly broken scale symmetry is essential for dense skyrmion matter in hidden local symmetry theory. Consistency with the vector manifestation fixed point for the hidden local symmetry of the lowest-lying vector mesons and the dilaton limit fixed point for scale symmetry in dense matter is found to require that the anomalous dimension ($\gamma_{G^2}$) of the gluon field strength tensor squared ($G^2$) that represents the quantum trace anomaly should be $0 <\gamma_{G^2} <3$.
\end{abstract}


\maketitle

\section{The Problem}
It is well-recognized and generally accepted in nuclear theory that two ingredients are essential in nuclear dynamics,  a large attraction and a large repulsion balancing each other providing the small nuclear binding energy observed in nature. In an effective field theory description, the former is simulated by the exchange of an iso-scalar scalar meson (denoted by $\sigma$) of mass $\sim 600$ MeV and the latter by that of an iso-scalar vector meson of mass {$\sim 780$ MeV} (identified with $\omega$).\footnote{In chiral perturbation theory anchored on nonlinear sigma model, those effects are simulated by high-order loop effects. The connection could be made but we won't dwell on it in this paper.} It is also accepted that when considered at the large $N_c$ limit in QCD, baryons can be described as skyrmions~\cite{Skyrme:1961vq,Witten:1979kh} and hence nuclear matter could be described as a skyrmion matter~\cite{Klebanov:1985qi}.

In this short note  we first point out that there is a serious problem in describing dense matter in terms of skyrmions due to a hitherto unobserved interplay between the scalar and vector degrees of freedom and then show that the problem can be resolved by resorting to an intricate interplay between scale-chiral symmetry and hidden local symmetry (HLS for short).

It was discovered in \cite{PRV1} that when skyrmions are put on lattice to simulate dense baryonic matter with a hidden gauge symmetric Lagrangian implemented with scale symmetry to take into account the change of vacuum in medium~\cite{BR91}, the energy of the system, the medium-modified pion decay constant $f_\pi^\ast$ and masses of the vector mesons and the pion diverge because of the divergence of the medium-modified expectation value of dilaton $\langle \chi\rangle^\ast$. This observation is in stark contrast to the intuitive expectation that, at very high density, chiral symmetry should be restored and hadronic phase should transit to quark(-gluon) phase. Moreover, in the HLS approach, it was found that at very high density or temperature the vector manifestation (VM for short)  fixed point should be approached with $f_\pi \to 0, m_\rho \to m_\pi = 0$~\cite{Harada:2000kb}. In addition, in the dilaton compensated chiral effective theory of baryonic matter, it was expected that at high density, the theory should go toward a dilaton-limit (DL) fixed point {characterized by $f_\sigma^\ast \to 0$}~\cite{beane,DLFP}. What's observed in \cite{PRV1} is therefore consistent with neither the {intuitive} expectation nor effective theory calculations.

The Lagrangian involved in \cite{PRV1} consisted of the standard (normal-parity) HLS Lagrangian with {an expicit} chiral symmetry breaking mass term, $O(p^2)$ in chiral counting (with $O(p^4)$ ignored), plus the anomalous, homogeneous Wess-Zumino (hWZ) term, $O(p^4)$ in chiral counting. The $\omega$ meson, crucial in nuclear dynamics, enters by coupling  to other degrees of freedom via the hWZ term, so the latter is essential for taking into account the $\omega$ meson in the dynamics even though it is {at the next-to-leading order in the chiral counting,} that is of $O(p^4)$. As has been well established, the skyrmion model with HLS Lagrangian with $\pi$ and $\rho$ -- and without $\omega$ -- fares well in finite nuclei as well as in nuclear matter, so the culprit in the trouble encountered in \cite{PRV1} was the $\omega$ associated with the hWZ term. Indeed the presence of the lowest iso-vector vector mesons $\rho$ and iso-vector axial-vector mesons $a_1$ improves the structure of finite nuclei over the Skyrme model (with pion only), with further improvement moving toward a BPS structure by an infinite tower~\cite{sutcliffe}.  The problem arises, however, when the  hWZ term is taken into account~\cite{Ma:2012kb}.

By itself, {the hWZ term} is classically scale-invariant, so the dilaton field $\chi$ does not figure in that term. Therefore,  the condensate $\la\chi\ra$ which controls the in-medium behavior of the Lagrangian~\cite{BR91} does not enter in the hWZ term in medium. The hWZ contribution to the energy of the skyrmion matter is found to diverge when integrated over the space unless the strong $\omega$ coupling to other degrees of freedom is suppressed. As a way-out, in \cite{PRV2}, a factor $(\chi/f_\sigma)^n$ was multiplied to the hWZ term,  {\it explicitly} breaking the scale invariance, and $n$ was varied  so as to tame the divergence. It was found that with $n\approx 2$ or 3, the problem was removed. However in \cite{PRV2},  this procedure was totally arbitrary, with justification from neither theory nor experiment. In this note we derive this result by generalizing the notion of scale-chiral effective theory proposed by Crewther and Tunstall~\cite{CT} {(CT for short)} to hidden local symmetry~\cite{LMR}.~\footnote{A similar idea was used by Yamawaki~\cite{yamawaki-GEB} in a somewhat different context in a dilatonic Higgs approach to beyond the Standard Model.}

\section{Hidden-Scale and Hidden-Local Symmetric Lagrangian}

We start with the HLS Lagrangian proposed in \cite{Bando:1984ej,HY:PR} implemented with scale symmetry that was used in \cite{PRV1}. It is in the form
\be
{\cal L}_{\rm HLS}& = & f_\pi^2\left(\frac{\chi}{f_\sigma}\right)^2{\rm
Tr}[\hat{a}_{\perp\mu}\hat{a}_{\perp}^{\mu}] + a f_\pi^2\left(\frac{\chi}{f_\sigma}\right)^2 {\rm Tr}[\hat{a}_{\parallel\mu}\hat{a}_{\parallel}^{\mu}] \nonumber\\
 & -&  \frac{1}{2g^2}{\rm Tr}[V_{\mu\nu}V^{\mu\nu}] + {\cal L}_{\rm hWZ}\nonumber\\
 &+& \frac 12 \partial_\mu\chi\partial^\mu\chi + {\cal V}(\chi), \label{prv1}
\ee
where the relevant degrees of freedom that figure are -- in addition to the pion field $U=e^{2i\pi/f_\pi}=\xi_L^\dag(x)\xi_R(x)$ -- the $U(2)$ vector fields $V_\mu=\frac 12 (\rho_\mu+\omega_\mu)$ and the conformal compensator field $\chi=f_\sigma e^{\sigma/f_\sigma}$ with $\sigma$ being the dilaton field.
In Lagrangian (\ref{prv1}), ${\cal L}_{\rm hWZ}$ is the homogeneous Wess-Zumino  term that consists of three independent terms that are reduced to one term in \cite{PRV1} as specified below and ${\cal V}(\chi)$ is the dilaton potential term that accounts for the trace anomaly of QCD. For simplicity, without the loss of generality in our discussion, we can take the chiral limit.

Several comments are in order before proceeding further. Apart from the dilaton potential ${\cal V}(\chi)$, the Lagrangian (\ref{prv1}) is scale invariant. It is of $O(p^2)$ in the chiral-scale counting except for ${\cal V}(\chi)$ and ${\cal L}_{\rm hWZ}$. We will explain later why $O(p^4)$ terms in the HLS sector will not affect our argument. The hWZ term is scale-dimension-4, hence scale invariant. But it is of $O(p^4)$ in the chiral-scale counting. It is this term that does the havoc to the dense skyrmion matter but cannot be ignored because it is the only way the $\omega$ meson enters in nuclear dynamics.

In order to resolve the disaster encountered in \cite{PRV1}, we resort to the strategy proposed by Crewther and Tunstall~\cite{CT} in implementing scale symmetry to chiral Lagrangian. How this strategy resolves a variety of scalar-meson conundrum in nuclear physics was discussed in \cite{LPR}. Here we follow CT formulated for non-linear sigma model and apply it to HLS. This is made feasible given that HLS is gauge-equivalent to non-linear sigma model~\cite{Bando:1984ej,HY:PR}. The dilaton implemented HLS, ${\cal L}_{{\rm HLS}_{\sigma}}$, generalized from CT considered in this note is given by~\cite{LMR}
\begin{eqnarray}
\cal L_{{\rm HLS}_\sigma}& = & {\cal L}_{{\rm HLS}_\sigma}^{d=4} + {\cal L}_{{\rm HLS}_\sigma}^{d > 4} + {\cal V}(\chi),\label{eq:CTHLS}
\end{eqnarray}
with
\begin{widetext}
\begin{subequations}
\begin{eqnarray}
{\cal L}_{{\rm HLS}_\sigma}^{d=4} & = & f_\pi^2h_1 \left(\frac{\chi}{f_\sigma}\right)^2{\rm
Tr}[\hat{a}_{\perp\mu}\hat{a}_{\perp}^{\mu}] + a f_\pi^2 h_2  \left(\frac{\chi}{f_\sigma}\right)^2{\rm
Tr}[\hat{a}_{\parallel\mu}\hat{a}_{\parallel}^{\mu}] -
\frac{1}{2g^2}h_3{\rm Tr}[V_{\mu\nu}V^{\mu\nu}] + \frac{1}{2}h_4 \partial_\mu \chi \partial^\mu \chi +
c_h \mathcal{L}_{{\rm hWZ}}, \label{eq:CTHLS40}\\
{\cal L}_{{\rm HLS}_\sigma}^{d > 4} & = & f_\pi^2(1-h_1) \left(\frac{\chi}{f_\sigma}\right)^{2+\beta^\prime}{\rm
Tr}[\hat{a}_{\perp\mu}\hat{a}_{\perp}^{\mu}] + (1-h_2) af_\pi^2\left(\frac{\chi}{f_\sigma}\right)^{2+\beta^\prime}{\rm
Tr}[\hat{a}_{\parallel\mu}\hat{a}_{\parallel}^{\mu}]\nonumber\\
& &{}  -
\frac{1}{2g^2}(1-h_3)\left(\frac{\chi}{f_\sigma}\right)^{\beta^\prime}{\rm Tr}[V_{\mu\nu}V^{\mu\nu}] + \frac{1}{2}(1-h_4) \left(\frac{\chi}{f_\sigma}\right)^{\beta^\prime} \partial_\mu \chi \partial^\mu \chi + (1-c_h)\left(\frac{\chi}{f_\sigma}\right)^{\beta^\prime} \mathcal{L}_{{\rm hWZ}}, \label{eq:CTHLSg40}\\
{\cal V}(\chi) & = &{}  h_5 \left(\frac{\chi}{f_\sigma}\right)^4 + h_6 \left(\frac{\chi}{f_\sigma}\right)^{4+\beta^\prime}.\label{eq:CTHLSdV}
\end{eqnarray}
\end{subequations}
\end{widetext}

In this Lagrangian that takes into account the quantum anomaly to the leading order in chiral-scale symmetry, the scale-invariant term ${\cal L}_{{\rm HLS}_\sigma}^{d=4}$ comes directly from the conformal compensated HLS Lagrangian while the scale-noninvariant term ${\cal L}_{{\rm HLS}_\sigma}^{d>4}$ arises from the explicit scale symmetry breaking accounted for by the slope of the beta function $\beta^\prime$ which is equal to the anomalous dimension of $\gamma_{G^2}$ at a (presumed) IR fixed point. Because of the slope of the beta function, or equivalently the anomalous dimension of the gluon $G^2$, $\gamma_{G^2}$, although the hWZ terms ${\cal L}_{{\rm hWZ}}$ is scale-invariant at the classical level, the dilaton couples to them due to the quantum anomaly.  As mentioned before, we ignore the $O(p^4)$ terms in the HLS sector.

In the  Lagrangian~\eqref{eq:CTHLS}, the low-energy constants $h_i (i = 1, \cdots, 6)$ cannot all be fixed unambiguously. Here, however, modulo explicit scale symmetry breaking which brings in additional $O(p^2)$ chiral-scale order \`a la Crewther and Tunstall~\cite{CT},  we can take $h_i = 1, (i =1,2,3, 4)$. As mentioned in the last section, this is justified as we are working with $O(p^2)$ normal-parity component of HLS.
With this choice, the Lagrangian~\eqref{eq:CTHLS} reduces to
\begin{widetext}
\be
{\cal L}_{{\rm HLS}_{\sigma}}= f_\pi^2 \left(\frac{\chi}{f_\sigma}\right)^2{\rm
Tr}[\hat{a}_{\perp\mu}\hat{a}_{\perp}^{\mu}] + a f_\pi^2  \left(\frac{\chi}{f_\sigma}\right)^2{\rm
Tr}[\hat{a}_{\parallel\mu}\hat{a}_{\parallel}^{\mu}] -
\frac{1}{2g^2}{\rm Tr}[V_{\mu\nu}V^{\mu\nu}] + \frac{1}{2} \partial_\mu \chi \partial^\mu \chi + \tilde{{\cal L}}_{\rm hWZ} + {\cal V}(\chi).
\label{leading}
\ee
\end{widetext}
As stated, we reduce the hWZ term to a single term which can be obtained by a suitable choice of the coefficients of the three terms,
\be
\tilde{{\cal L}}_{\rm hWZ} = g\omega_\mu B^\mu \left(c_{h}+(1-c_{h})\left(\frac{\chi}{f_\sigma}\right)^{\beta^\prime}\right) \label{hWZ}
\ee
where $B_\mu$ is the topological (baryon) current.
The constant $c_h$ is an unknown constant together with the anomalous dimension $\beta^\prime$. One can actually treat all three hWZ terms as was done in \cite{Ma:2013ela}. For our purpose this is unnecessary. We have assured ourselves that the qualitative feature we are treating is not affected.

Fluctuating the dilaton field with respect to its vacuum value and keeping the leading-order fluctuation, one can express the dilaton potential as
\begin{eqnarray}
{\cal V}(\chi) & = & {} \frac{m_\sigma^2 f_\sigma^2}{4+\beta^\prime} \left( \frac{\chi}{f_\sigma}\right)^4 \left[\ln \left( \frac{\chi}{f_\sigma}\right) - \frac{1}{4}\right].\label{potential}
\label{eq:potentiallog}
\end{eqnarray}
To obtain this expression we have used that, in the chiral limit, $m_\sigma^2f_\sigma^2 = - 4 \beta^\prime h_5 = \beta^\prime(4+\beta^\prime)h_6$ which is the dilaton analog to the Gell-Mann-Oakes-Renner relation for the pion with $m_\pi^2\sim O(p^2)$~\cite{CT}. If $\beta^\prime$ were $\ll 1$, at the leading order of $\beta^\prime$, the dilaton potential would go to the log-type potential widely used in the literature~\cite{Goldberger:2008zz}. But here we will keep it as a parameter since it should be of $O(1)$ to realize the VM and the dilaton-limit as we will see later.

\section{The Solution}

The solution to the problem encountered in \cite{PRV1} turns out to be amazingly simple.  The solution given in \cite{PRV2} corresponds to (\ref{hWZ}) with
\be
2\lsim \beta^\prime\lsim 3, \ \  c_h\approx 0\label{prv2}
\ee
with no other modifications. There the same Lagrangian as (\ref{leading}) and the potential (\ref{potential}) were used but the factor $(\chi/f_\sigma)^n$ with $n\lsim 3$ was multiplied without any justification to  $g\omega_\mu B^\mu$ to simply quench the gauge coupling $g$ for increasing density. Here, as explained in the last section, the QCD trace anomaly is seen to do the job.

It is extremely interesting to explore what information one can get on the role of explicit scale symmetry breaking characterized by $\beta^\prime\neq 0$ from dense baryonic matter. There is no information that can be trusted on $\beta^\prime$ in QCD with $N_c=3$ and $N_f < 8$ from experiments or theory. We are aware of calculation of $\beta^\prime$ in scheme-independent series expansion for $N_f >9$ but it does not give any hint on QCD proper with three flavors. Therefore it is intriguing that dense nuclear matter may be able to say something, tentative though it may be, about the anomalous dimension {of the gluon field strength tensor squared}.

For this purpose, we proceed to look at the model defined above and simulate dense skyrmion matter with two parameters $\beta^\prime$ and $c_h$. To gain a rough idea,  we take the parameters used in \cite{CT}, $m_\sigma = 550~$MeV and $f_\sigma = 100$~MeV. Other values in the vicinity of these would not affect the results qualitatively.

To confront nature, i.e., the EoS for compact stars, the $U(2)$ symmetry for the vector mesons $V=(\rho,\omega)$ that holds well in the matter-free vacuum cannot remain unbroken at density $n\gsim 2n_0$~\cite{PKLR,PR2016}. With breaking of $U(2)$ symmetry for the vector mesons, the half-skyrmion phase which controls the symmetry energy in compact-star matter sets in at {$n\sim 2n_0$~\cite{Ma:2013ela}}. This structure is not present with the $U(2)$ symmetry for the $V_\mu$ and the low-mass  dilaton, $m_\sigma = 550~$MeV. Both of these features are not realistic since the $U(2)$ symmetry should be strongly broken in dense matter~\cite{PKLR} and the dilaton mass could be higher in the matter-free vacuum. However the qualitative feature we are concerned with is the divergence of the decay constant $f_\sigma\approx f_\pi$ and the consequent violent deviation from  the concordant VM/DL fixed point, and this behavior is indifferent to the presence of the half-skyrmion phase.

In the model so defined, the density dependent $f_\pi^\ast, m_\rho^\ast, m_\omega^\ast$ and $f_\sigma^\ast$ can be expressed as~\cite{PRV1,Ma:2013ela}
\begin{eqnarray}
\frac{f_\pi^\ast}{f_\pi} & \approx  & \sqrt{\left\langle \left(\frac{\chi}{f_\sigma}\right)^2\left[1 - \frac{2}{3}\left(1 - {\frac{1}{4}\left({\rm Tr}U\right)^2}\right)\right]\right\rangle}\, , \nonumber\\
\frac{f_\sigma^\ast}{f_\sigma} & \approx & \frac{m_\rho^\ast}{m_\rho} = \frac{m_\omega^\ast}{m_\omega} = \sqrt{\left\langle\left(\frac{\chi}{f_\sigma}\right)^2\right\rangle} \, .
\end{eqnarray}
The possible values of $c_h$ and $\beta^\prime$ to avoid the divergence problem and yield VM and dilaton limit fixed point at high density are summarized in Table~\ref{table:values} marked with "$\surd$." Those indicated by "$\times$" are excluded because the VM and DLFP set in at much too low  densities. From this table, one can conclude that, to be consistent with the VM and DL fixed points, the ranges of values acceptable are $0 < \beta^\prime < 3.0$ and $0\leq c_h \leq 0.3$. It seems safe to say that $\beta^\prime\approx 0$ is ruled out from what's found in dense matter. These ranges could be sharpened if the unrealistic features mentioned above were suitably accounted for.

\begin{table}[t]
\caption{\label{table:values} Possible values of $c_h$ and $\beta^\prime$ to yield VM and DLFP.}
\centering
\begin{tabular}{c|llllll}
\hline
\backslashbox{$c_h$}{$\beta^\prime$}& \quad $1.0$ &\quad $1.5$ & \quad $2.0$ & \quad $2.5$ & \quad $3.0$ & \quad $3.5$ \\
\hline
$0.0$ &\quad $\surd$ &\quad $ \surd $ &\quad $ \surd $ &\quad $ \surd $ &\quad $ \times $ &\quad $ \times $ \\
\hline
$0.1$ &\quad $\surd$ &\quad $\surd $ &\quad $\surd$ &\quad $\surd$ &\quad $\surd$ &\quad $\times$ \\
\hline
$0.2$ &\quad $\surd$ &\quad $ \surd$ &\quad $\surd$ &\quad $\surd$ &\quad $\surd$ &\quad $\times$ \\
\hline
$0.3$ &\quad $\surd$ &\quad $\surd$ &\quad $\surd$ &\quad $\surd$ &\quad $\surd$ &\quad $\surd$ \\
\hline
\end{tabular}
\end{table}

\begin{figure}[H]\centering
\includegraphics[scale=0.14]{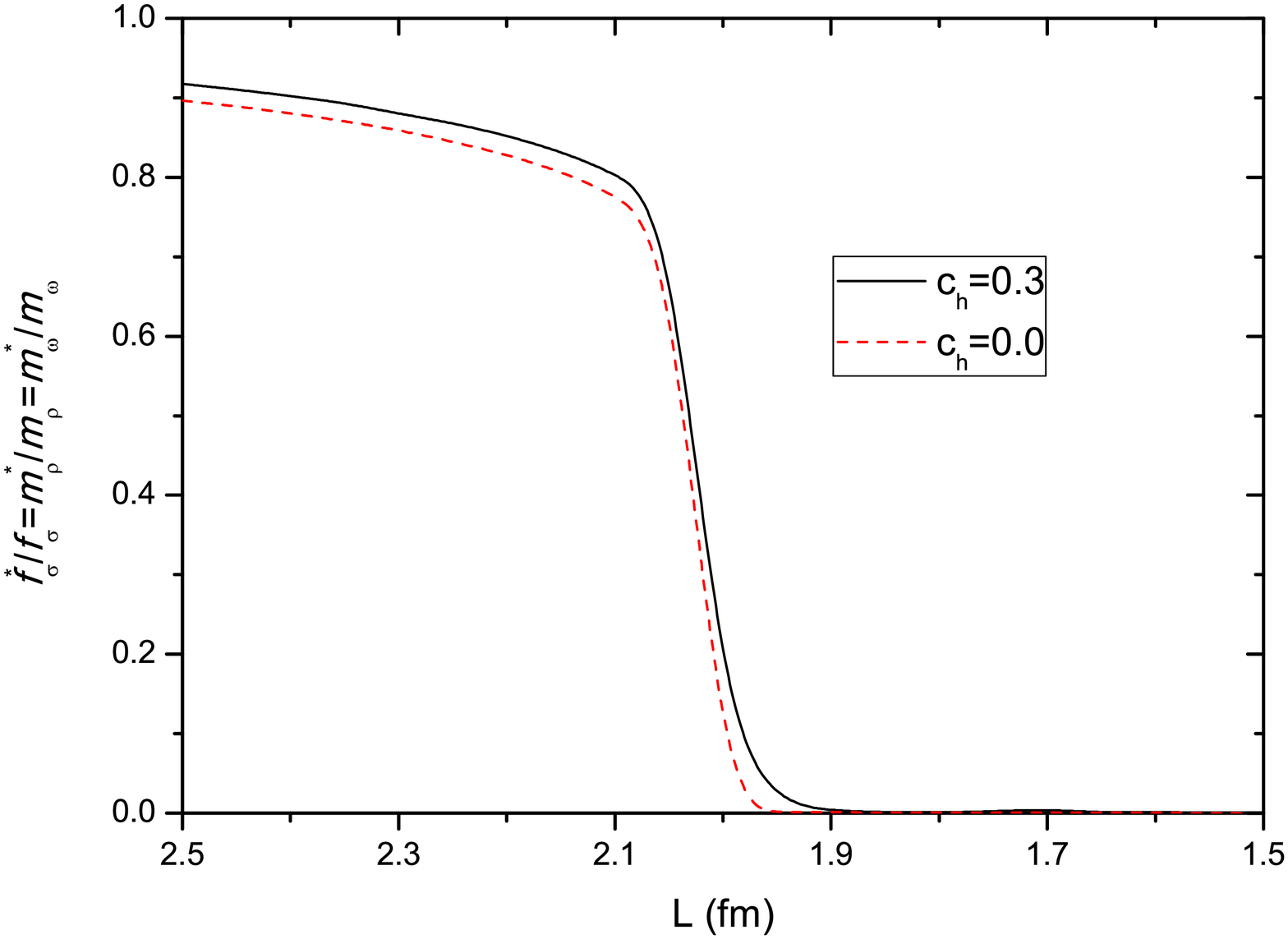}\includegraphics[scale=0.14]{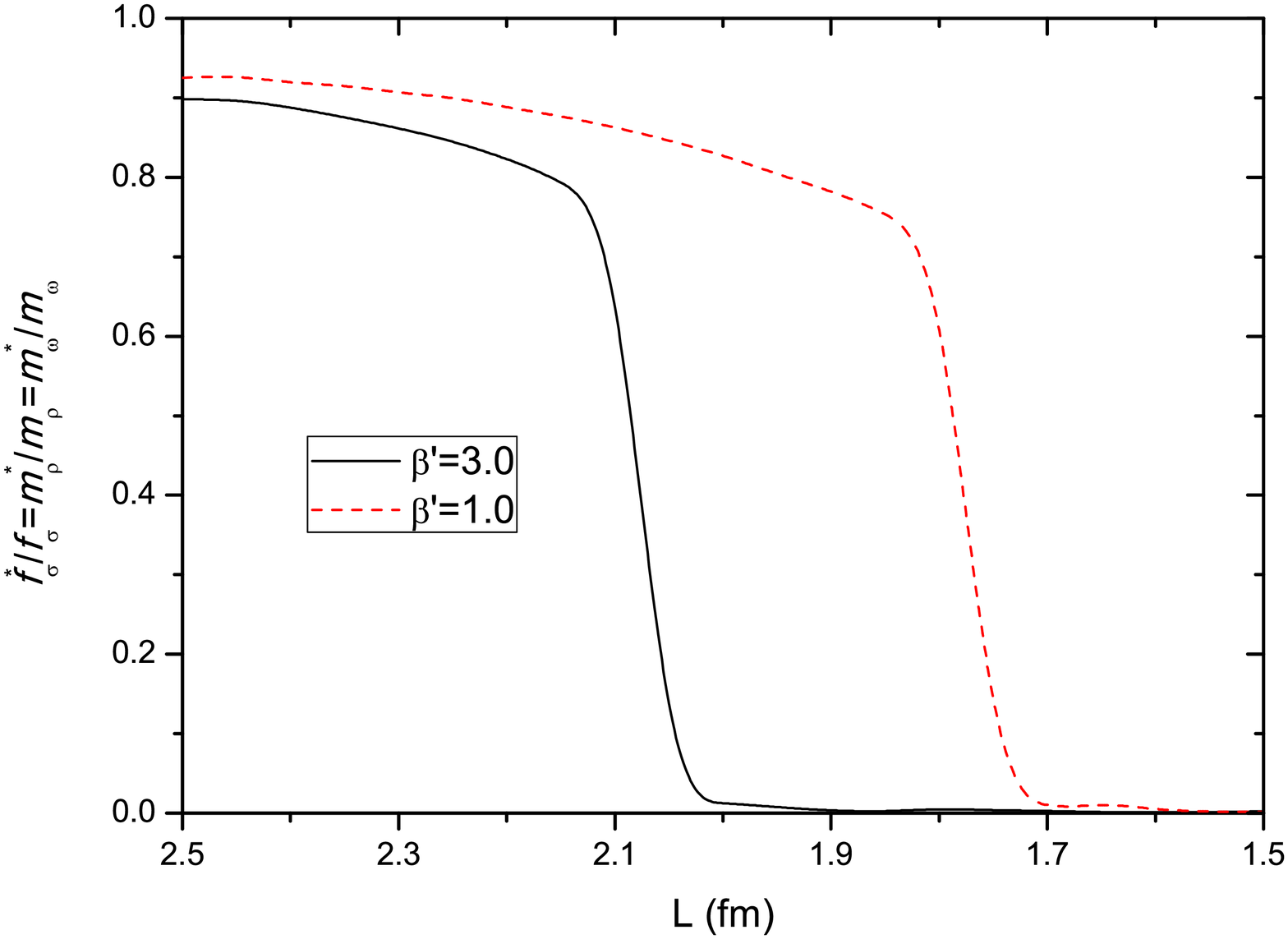}
\caption{The effects of $c_h$ (left panel for $\beta^\prime = 2.5$) and $\beta^\prime$ (right panel for $c_h = 0.2$) on the hadron properties as a function of crystal size $L$. The similar plots hold for $f_\pi^\ast/f_\pi$. }
\label{fig:Plot}
\end{figure}

We next show the effect of the parameters $c_h$ and $\beta^\prime$ on hadron properties in Fig.~\ref{fig:Plot}. This figure shows that the hadron properties are not significantly changed by $c_h$ as long as it is $\lsim 0.3$. However, a smaller $\beta^\prime$, even though it avoids the divergence, pushes chiral restoration to an unacceptably high density.

\section{The Implications}

Several intriguing predictions follow from the above observation.

The first is that in dense matter, the $\omega$ coupling to other degrees of freedom is suppressed by the explicit breaking of scale symmetry.  This would also ameliorate a similar disturbing situation in holographic models. For instance in the Sakai-Sugimoto model~\cite{sakai-sugimoto}, at the leading order  $O(N_c\lambda)$ where $\lambda$ is the t'Hooft constant, the 5D YM Lagrangian in flat space describes baryonic matter as a BPS-instanton matter. It has zero binding energy, which is close to nature where the binding energy of heavy nuclei is tiny. However this picture breaks down at the next order in $\lambda$, i.e., at $O(N_c\lambda^0)$. At that order,  the Chern-Simons term with $U(1)$ vector field  -- which corresponds to the $\omega$ meson in 4D --  enters and  spoils drastically the nice agreement obtained at the leading order~\cite{Ma-Rho-review}.\footnote{At this order the space warping also contributes but this effect, though less well controlled, is expected to be less important.}  This  can also be seen in  the bottom-up approach of Sutcliffe~\cite{sutcliffe} where it is found that the ``good result"  obtained by adding more and more iso-vector and axial-vector mesons in the skyrmion structure gets spoiled when the $\omega$ fields are taken into account.  The result we obtained here suggests that this problem can also be resolved by the $\beta^\prime$ effect when the broken scale symmetry is suitably implemented into the Chern-Simons term.

A potentially more significant implication of this result is that dense baryonic matter can provide information on the fundamental quantity, the anomalous dimension $\beta^\prime$, and the constant $c_h$. There is a recent work which purports to access this quantity in gauge theories, but for $N_f > 9$~\cite{shrock}\footnote{The scheme-independent series expansion performed in this paper --  not done for  $N_f < 9$ --  indicates that the upper limit is likely more relevant.}.  So this quantity -- as well as $c_h$ -- is essentially unknown. What one gets in dense matter is therefore quite suggestive and deserves further studies.

There is an intriguing hint that HLS limited to $O(p^2)$ gives highly non-trivial predictions that are unavailable from other approaches~\cite{HY:PR,komargodski}. For instance, the KSRF relation, which works surprisingly well, is found to hold to {\it all} loop orders~\cite{HY:PR}.  There is also the vector dominance phenomenon which is otherwise hard to derive. All this put together would suggest that in scale-invariant HLS applied to dense matter, one may ignore $O(p^4)$ terms except for  the hWZ terms -- which are essential. A numerical evidence for this is given in \cite{Ma-Rho-review}. In medium, the same suppression as is observed in the hWZ terms could be effective for {\it all}  $O(p^4)$ terms, in which case the leading order Lagrangians given above should be reliable, making the theory much simpler than thought.

\subsection*{Acknowlegments}

One of the authors (MR) is grateful to Rod Crewther and Lewis Tunstall for helpful correspondence. The work of Y.~L. Ma was supported in part by National Science
Foundation of China (NSFC) under Grant No. 11475071, 11547308 and the Seeds Funding of Jilin
University.


\end{document}